\documentclass[]{spie}  

\usepackage{amsmath,amsfonts,amssymb}
\usepackage{graphicx}
\usepackage[colorlinks=true, allcolors=blue]{hyperref}

\usepackage{graphicx}
\usepackage{booktabs}
\usepackage[justification=centering]{caption}
\usepackage{listings}
\usepackage{color}
\usepackage{soul}
\usepackage{float}
\usepackage{subcaption}

\definecolor{dkgreen}{rgb}{0,0.6,0}
\definecolor{gray}{rgb}{0.5,0.5,0.5}
\definecolor{mauve}{rgb}{0.58,0,0.82}

\title{ptychopy: GPU framework for ptychographic data analysis}

\author[a]{Ke Yue}
\author[a]{Junjing Deng}
\author[a]{Yi Jiang}
\author[a]{Youssef Nashed}
\author[a]{David Vine}
\author[a]{Stefan Vogt}

\affil[a]{Advanced Photon Source, Argonne National Laboratory, 9700 S Cass Ave, Argonne IL 60439 USA}

\authorinfo{Further author information: (Send correspondence to Ke Yue.)\\Ke Yue.: E-mail: kyue@anl.gov}

\begin{document}                  
\maketitle                        







\begin{abstract}
X-ray ptychography imaging at synchrotron facilities like the Advanced Photon Source (APS) involves controlling instrument hardwares to collect a set of diffraction
patterns from overlapping coherent illumination spots on extended samples, managing data storage, reconstructing ptychographic images from acquired diffraction patterns, and providing the visualization of results and feedback. In addition to the complicated workflow, ptychography instrument could produce up to several TB's of data per second that is needed to be processed in real time. This brings up the need to develop a high performance, robust and user friendly processing software package for ptychographic data analysis.  In this paper we present a software framework which provides functionality of visualization, work flow control, and data reconstruction. To accelerate the computation and large datasets process, the data reconstruction part is implemented with three algorithms, ePIE~\cite{maiden2009improved}, DM~\cite{thibault2009probe} and LSQML~\cite{thibault2013reconstructing} using CUDA-C on GPU.
\end{abstract}


 
\section{Introduction}

Beamlines at synchrotron facilities use X-rays, various instruments, and detectors to conduct a series of experimental scans for different purposes, such as fluorescence mapping, tomography, coherent diffraction imaging (CDI), Laue depth reconstruction, etc. Among these categories, CDI is a novel lensless technique which uses a coherent beam to reveal the structural information of a specimen of interest~\cite{miao1999}. By recording the diffraction pattern of the coherent beam diffracted from the sample, the sample's complex-valued transmission function can be reconstructed via a ``computation lens'' -- iterative phase retrieval algorithms. 

Ptychography, originally introduced by Hoppe in electron microscopy~\cite{hoppe1969beugung}, combines the advantages of scanning transmission X-ray microscopy (STXM) with CDI to provide high spatial resolution images on an extended object~\cite{faulkner2004movable,rodenburg2007,thibault2008high}. This new imaging technique is limited neither by the X-ray optics nor by the requirements of isolated samples in CDI. Ptychography involves recording a set of diffraction patterns by scanning a coherent probe across an extended sample with adjacent illumination spots overlapped. Redundant information among those significantly overlapped diffraction patterns is used to reconstruct the object transmission function and probe function in a phase retrieval algorithm. 

There are many phase retrieval methods for ptychographic reconstruction, such as relaxed averaged alternating reflections (RAAR) algorithm~\cite{luke2004relaxed}, difference map (DM) algorithm~\cite{thibault2009probe,thibault2008high}, hybrid input-output (HIO) algorithm by Fienup~\cite{fienup1982phase},  maximum likelihood (ML) principles~\cite{thibault2012maximum}, ptychographical iterative engine (PIE)~\cite{rodenburg2004phase}, and its extended version ePIE~\cite{maiden2009improved}. These methods try to recover the lost phase from a set of recorded diffraction patterns and provide different accuracy under certain circumstances. In addition to the accuracy of the reconstruction methods, the reconstruction speed is another important factor in ptychographic imaging. Due to the developments of brighter synchrotron sources, a highly nanofocusing beam combined with highly efficient scanning, and the higher throughput of fast detectors, the size of the resulting ptychographic datasets is increasing rapidly. Parallel computation on Graphics Processing Units (GPUs) has shown that it can accelerate data processing, especially beamline data analysis problem~\cite{pharr2005gpu,yue2015accelerating}, which usually have huge amount of data volume. GPU processing has been introduced for ptychography reconstructions to speed up phase retrieval, such as SHARP (Scalable Hetereogeneous Adaptive Real-time Ptychography)~\cite{marchesini2016sharp} and our previous work ptycholib~\cite{nashed2014parallel}. In ptycholib, the ePIE phase retrieval algorithm is implemented with Compute Unified Device Architecture (CUDA) to get a speedup factor of about two orders of magnitude on one GPU. To continue to accelerate the computation and  larger datasets process, a hybrid parallel strategy to divide the computation between multiple GPUs has been implemented to enable real-time ptychographic phase retrieval reconstruction. 

In addition to the reconstruction code, a complete ptychography experiment usually involves scan motion controls, data acquisition, metadata logging of experimental parameters, detector data transfer, data processing and reconstruction, and/or providing feedback. To enable high-throughput pytchography and user-friendly control for beamline users, many synchrotron facilities have been driven to develop software frameworks for high-rate data analysis. PtyPy~\cite{enders2016computational} is a python open-source software framework featuring clear separation between representations of physical experiments, the models, and the algorithmic implementations to solve the inverse problem. The cSAXS beamline at the Swiss Light Source develops some MATLAB software ~\cite{thibault2008high,thibault2012maximum,odstrcil_oe_18} implementing the DM and Maximum-likelihood algorithms, and a generalized data handling and reconstruction package is under development. The Advanced Light Source (ALS) developed a real-time streaming framework called Nanosurveyor~\cite{daurer2017nanosurveyor} which provides a streamlined processing pipeline to support distributed real-time analysis and visualization. The software's backend is based on SHARP which uses RAAR as the reconstruction algorithm and runs in parallel on multiple GPUs on a remote computer cluster.

At the APS, we are able to implement ptychographic data acquisition at a high frequency up to 3 kHz with a Dectris Eiger 500K detector~\cite{radicci_joi_2012} in continuous flyscan mode ~\cite{deng2015continuous,pelz_apl_2014,huang_scirep_2015, deng2019velociprobe}, yielding 6 GB/second raw data from the detector. After the APS upgrade project, the data rates from the detector could be up to several TB's per second. Setting up a streamlined workflow from the data collection instrument to the data processing unit and getting quick feedback is very crucial for high-throughput ptychography experiments. Our current reconstruction code, ptycholib, is implemented with algorithm ePIE using CUDA and C++, which is not easy to integrate with other modules in a python environment. For example, several synchrotron facilities use EPCIS (Experiment Physics and Industrial Control System) to control experiments, and there is a python module called PyEpics~\cite{newville2014pyepics} providing python interface to EPICS so that we can use python scripts to control the beamline instrument and collect the data. In addition, there are a lot of data analysis tools written in python  in the data processing and optimization area, which would also be needed in the ptychographic reconstruction workflow. Furthermore, the reconstruction is launched via a command line interface in a terminal, which is not user friendly for users who have little command line interface experience. To solve the above issue, an open-sourced python-based software framework called ptychopy ~\cite{ptychopy} is developed which could integrate the GPU-based reconstruction module with other beamline software module such as beamline control, data collection and storage modules. The data analysis part is implemented with ePIE ~\cite{nashed2014parallel}, DM~\cite{thibault2009probe,thibault2008high} and LSQML~\cite{thibault2013reconstructing, odstrcil_oe_18} using CUDA and wrapped with python API interface. Since the computation is still running on GPUs with CUDA through shared libraries, the performance in reconstruction speed does not degrade using our python package ptychopy. Within this framework, a frontend module is implemented with PyQt5 for visualization and parameter configuration, which interfaces with other modules to manage reconstruction and resultant storage, converts the image format from CSV to other formats such as TIFF, provides feedback to the user about the reconstruction status, and communicates results using log files and email. The ePIE, DM or LSQML phase retrieval algorithm with high-performance computing is running on the backend on a remote cluster. During the reconstruction, the status can be shown in real time in the frontend GUI, and the reconstructed results at the remote cluster could be visualized from the GUI. The software framework is running on the GPU and integrated with other beamline control and software packages for streamlined high performance data analysis. 


\section{Ptychopy software framework overview}

This section will describe the overview architecture of the software framework ptychopy, which includes three major components.

\subsection{Software architecture overview}

The architecture of the software is based on the model-view-controller (MVC) model, which includes three components, a user view interface for presenting and accepting information, a backend module for processing the information, and a controller component to manage the data, rules and logic to connect the interface to the backend module, as shown in Fig.~\ref{fig:gui}. The user view interface and controller component are implemented using python and the backend modules are implemented with CPython, C++, and CUDA. It provides python APIs for doing ptychographic reconstruction using three algorithms ePIE, DM, LSQML on GPU.

\subsection{Graphical User Interface}

 
The frontend component is implemented with the PyQt5 graphics toolkit for the GUI and visualization functionality. The major functionalities include providing input for the reconstruction parameters and visualizing the reconstruction results (see Fig.~\ref{fig:gui}). For the reconstruction parameters, there are four categories, which are:  

 \begin{figure}
 	\centering
 	\fbox{ \includegraphics[width=0.96\textwidth]{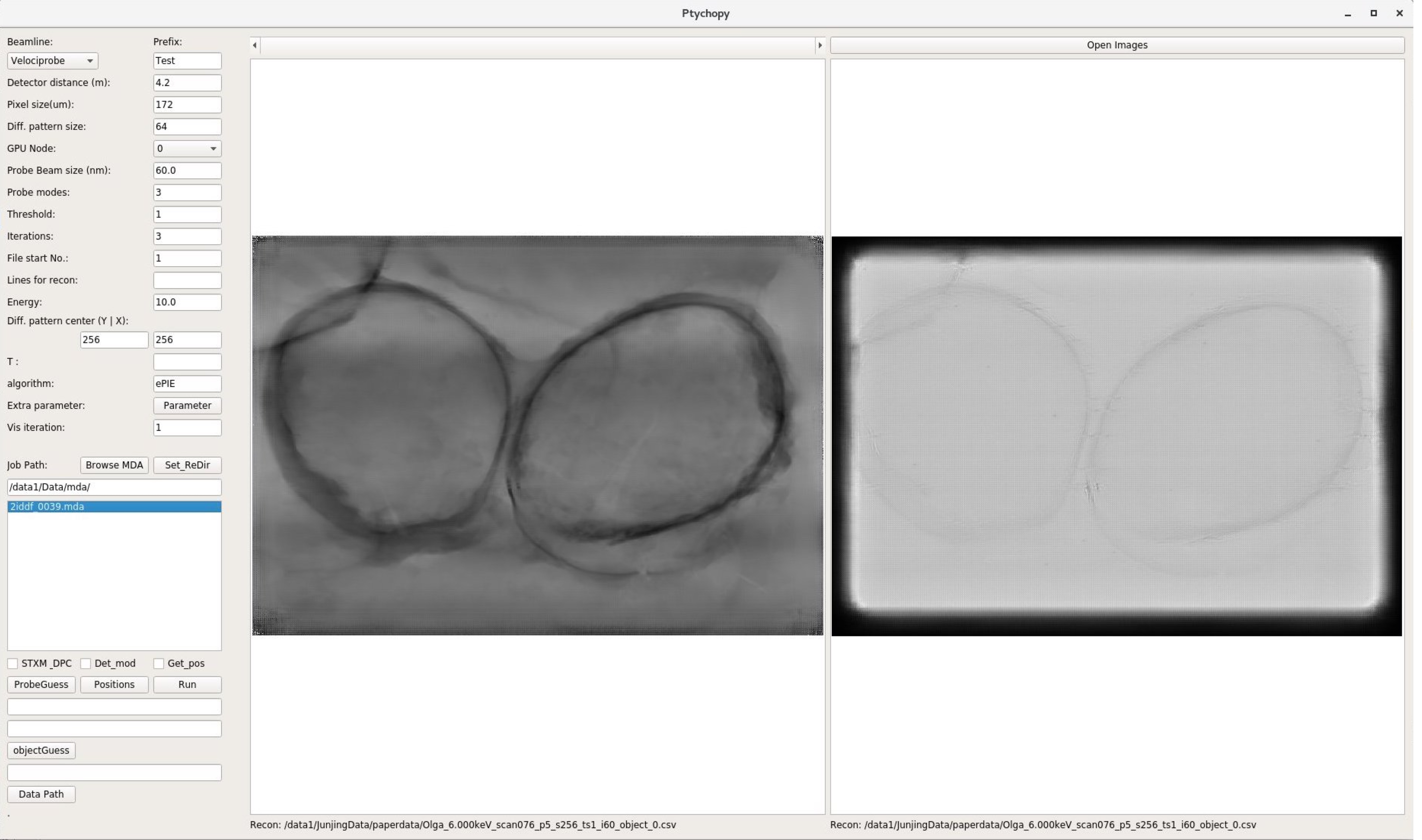}} 
 	\caption{Graphical User Interface for ptychopy  showing a ptychography reconstruction result with ePIE algorithm}
 	\label{fig:gui}
 \end{figure}

\begin{itemize}
\item Provide the interface for the experimental parameters such as beam energy, the guess of probe size, the number of probe modes, detector distance, scan dimensions and step size, and axis for flipping the scan direction. 

\item Define the parameters for preprocessing diffraction data, such as the center of the diffraction patterns, the cropping array size, the pixel bit-depth for detecting saturated pixels, the threshold value to remove the noise, the rotation angles of the diffraction patterns, and the number of diffraction patterns per data file.

\item Specify the parameters for the multiple GPU configuration, such as the GPU node index, the overlapping size where the data is shared between multiple GPUs, the share-frequency of data shared among multi-GPUs.

\item Set up the parameters for phase retrieval, such as the reconstruction algorithm (currently ePIE, DM and LSQML are available), and the number of iterations for phase retrieval, after which iteration to update the probe. The probe and/or object guess can also be specified in the GUI if they are available.
\end{itemize}

Most of the parameters for experiment scans and diffraction patterns are saved in MDA files by EPICS, which then can be directly imported into our GUI interface so that the users do not have to enter the parameters one by one. The editable text boxes on the GUI also allow users to tune the parameters to do customized reconstructions. 

 
\subsection{Data reconstruction back end module}

The ePIE, DM and LSQML algorithms use a set of far-field diffraction pattern data to reconstruct both the probe function and the object function simultaneously. The iterative procedure involves the computation of multiple Fast Fourier Transforms (FFTs) over thousands of diffraction patterns, making it suitable for parallel analysis on GPU. Compared to DM, ePIE needs a smaller memory footprint than DM, since DM has to keep a copy of the previous calculation in memory and calculate the exit wavefronts of all the scan points at once. But DM would run faster than ePIE on the reconstruction since DM can parallelize the calculation of the scan points at the same time on GPUs. LSQML's memory footprint requirement is between ePIE and DM while taking most time among three algorithms to produce the best quality of the images. Depending on the requirement on the reconstruction speed and the quality of the resultant image, the user could choose ePIE, DM or LSQML algorithm based on their own need.

The size of the ptychography datasets is increasing rapidly due to the developments of highly efficient scanning and fast detectors, which brings up the need of massive computation power. Homogeneous many-core processors with massively parallel structure, such as GPUs, are emerging in broad application areas and has been providing massive computation power for processing large blocks of data in parallel. CUDA is a parallel computing platform and programming model created by NVIDIA, which gives the developers access to the virtual instruction set and memory of the parallel computational elements in GPUs via a C programming environment. GPU programs using CUDA have been widely implemented on many scientific problems and have been improving the program performance dramatically. Therefore, in the ptychopy reconstruction module, we implemented the algorithms with C and CUDA as the backend for the performance enhancement purposes.

The backend module is a CPython module which provides ePIE, DM, and LSQML reconstruction algorithm options for phase retrieval. All the reconstruction algorithms are implemented using CUDA and C++ for performance consideration. CPython is a python interpreter implemented with C, providing the python interface for the CUDA code and also serves as the glue for the data reconstruction backend and other modules that are implemented in a python environment.

\subsection{Controller component}

The controller component uses threads for pipelining the reconstruction procedure and visualization procedure. The main thread is responsible for accepting the parameters from the GUI and reading the data from the HDF5 file. The worker thread is calling the backend reconstruction module and saving the results as CSV files. In addition, the controller component handles all the data saving, loading, and pipelining between the GUI and the reconstruction. 


\section{Reconstruction workflow}
In this section, the reconstruction workflow will be described in detail starting from data preprocessing, to data analysis and result feedback.

 \begin{figure}
 	\centering
 	\fbox{ \includegraphics[width=0.96\textwidth]{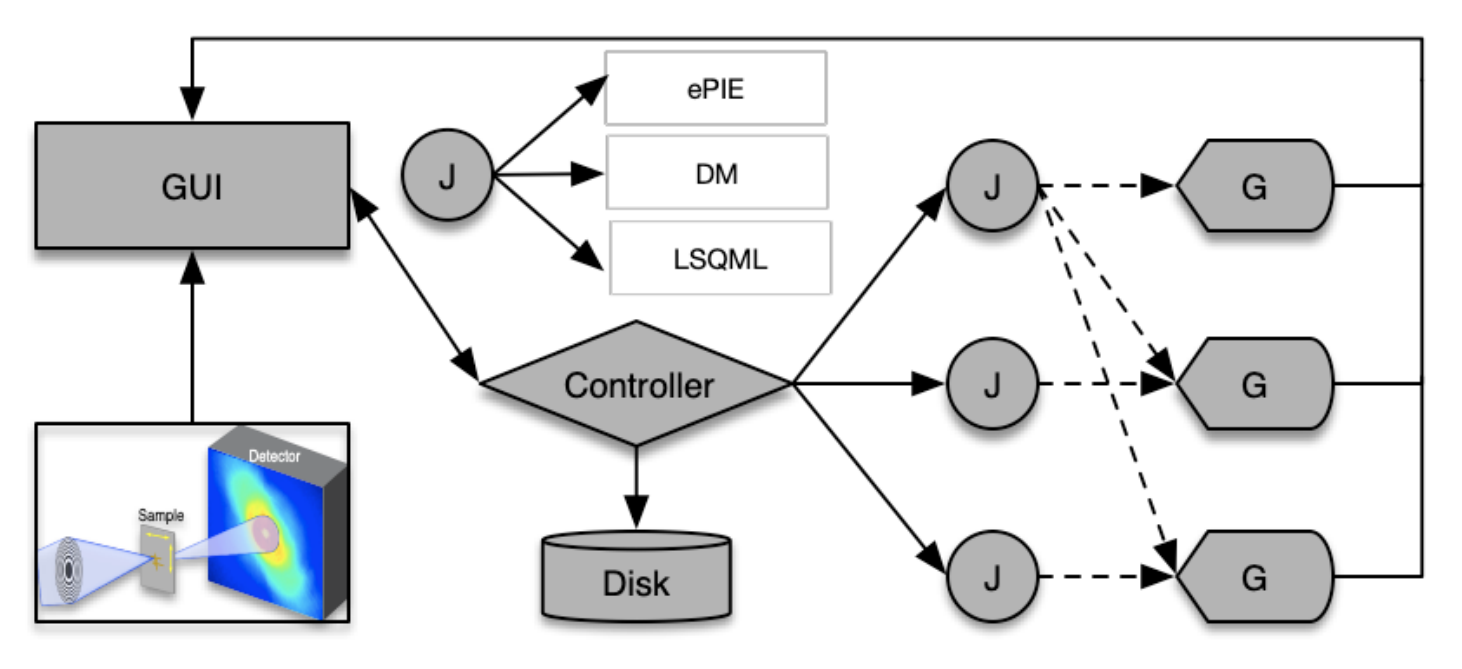}} 
 	\caption{Work flow for the beamline ptychographic reconstruction where J represents each reconstruction job and G represents the GPU }
 	\label{fig:flow}
 \end{figure}

\subsection{Synchron data parameter preprocessing}
The data collected at APS synchrotron beamlines is saved in HDF5 format and the parameters for the ptychography scan (such as energy, scan dimensions, step size, and rotation angle) and detector parameters are saved in the scan MDA file. ptychopy will read these parameters as default from the MDA file via its controller module. These default parameters can also be changed with customized ones if they are specified in the GUI. For batch reconstruction, the MDA files of the selected scans can be loaded as a list and the controller will read the MDA file list and corresponding parameters in a queue. In addition, to reduce the collected data size, the HDF5 datasets are compressed (by LZ4 or zlib) during data collection, which are automatically uncompressed when the data is imported for reconstructions. The collected data are usually a set of diffraction patterns and each HDF5 saves certain number of diffraction patterns. At APS beamlines, we save each row of the scan diffraction patterns into a separate file and use the row number as the HDF5 filename suffix. Our data analysis module of ptychopy will read the diffraction patterns from the HDF5 with the corresponding row and column numbers as the scan positions into GPU memory. The number of rows and columns of the scan positions can be defined using scanDims parameters showed in the following script example.


\subsection{GUI mode and script mode}

Ptychopy provides flexibility in that it can be used in either GUI or script mode. The interface of the GUI mode shown in Fig.~\ref{fig:gui} conveniently allows users who have little command line interface experience to specify the scan and reconstruction parameters by choosing the scan MDA files and/or editing those parameters in the GUI. If the parameters are not specified, the default value saved in the MDA file will be used for the reconstruction. After the reconstruction, the final result will be saved in CSV files, and meanwhile, the reconstructed results will be shown on the visualization window. The reconstruction results can also be viewed anytime using the image browser window on the GUI. 

\begin{figure}
 	\centering
 	\fbox{ \includegraphics[width=0.96\textwidth]{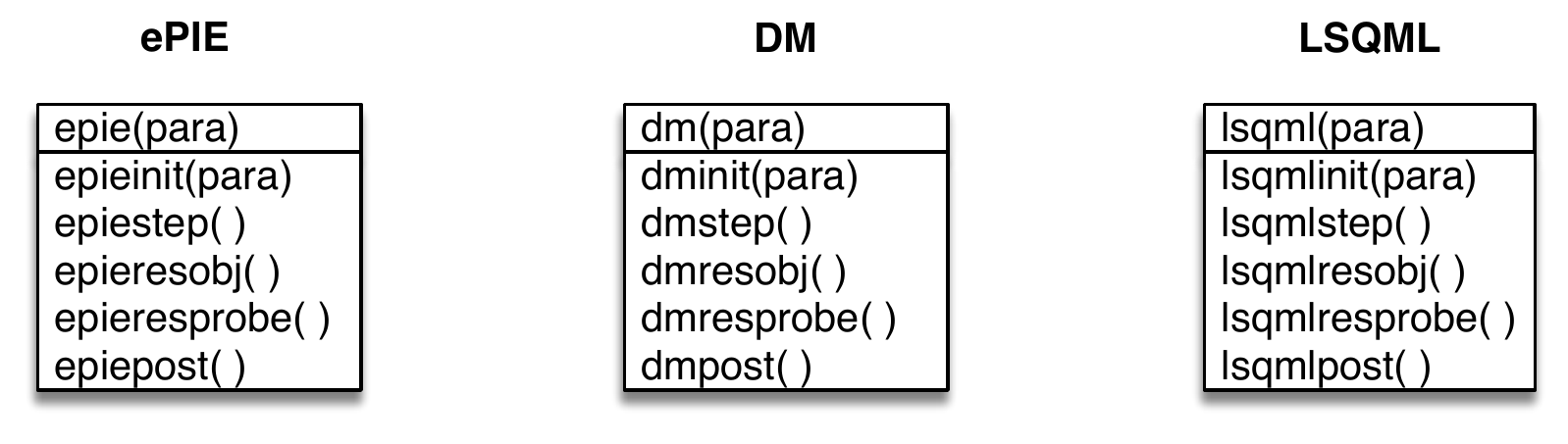}} 
 	\caption{Ptychopy APIs for ePIE, DM, LSQML methods  }
 	\label{fig:function}
 \end{figure}

In the script mode, the module may be integrated with beamline controls (such as PyEpics) and data management packages to achieve streamlined data collection and processing. The APIs for ePIE, DM and LSQML are illustrated in the Fig.~\ref{fig:function} for running as the whole mode or step mode. For whole mode, the reconstruction will run as a whole without interruption after passing parameters as para to the function. For the step mode, the result during the reconstruction could  be evaluated for a specified number of iterations. Taking ePIE as the example, para in the function epie(para) indicates the reconstruction parameters which are passed to the function and epie(para) will handle all the reconstruction on the GPU. If the value needs to be checked during the reconstruction, the step APIs in the Fig.~\ref{fig:function} could be used. Since the reconstruction algorithm is running on the GPU, retrieving the result from GPU needs to transfer the value back to CPU side and will take more time than running ePIE as a whole. If performance matters most, the whole  mode should be used instead of the step mode for running the algorithm.

Listing 1 shows a script example using the ePIE algorithm with a simulated image about how to use ptychopy module in the python script. The simulated images and test scripts are found in the source packages.  As shown in this example, the module accepts a string with the reconstruction parameters and returns the reconstruction results. The parameters can be found correspondingly in the GUI as a hint text box when one moves the cursor over a GUI component.

The first step is to initialize the reconstruction process and GPU device status with the epieinit() function. This includes initializing reconstruction parameters, IO processing, data loading and copying, initializing sample and probe array, copying the data from CPU side to GPU side, and preprocessing the data, such as diffraction pattern normalization. Then the number of iterations is defined and the ePIE reconstruction is launched with the epiestep() function. The phase retrieval procedure is running on the GPU side for a specified number of iterations. Before the reconstruction is done, the updated object and probe functions still reside on the GPU. To get the object and probe results, epieresobj() and epieresprobe() can be called anytime to return the data as a numpy array. Therefore, for the live view of the reconstruction process the updated probe and object results can be displayed at a defined iteration frequency from the GPU side. After the reconstruction, epiepost() function should be called to clean up the whole process and GPU device.

\begin{lstlisting}[language=python, caption=ptychopy script example using simulation image with ePIE algorithm]
import ptychopy

# Define the command string with parameters for reconstruction
cmdstr="./ptycho -jobID=sim512 -beamSize=110e-9 -scanDims=30,30 -step=50e-9,50e-9 -i=100 -size=512 -lambda=2.4796837508399954e-10 -dx_d=172e-6 -z=1 -simulate=1 -blind=0"

# Define the number of iterations
its = 100

# Init the reconstruction
ptychopy.epieinit(cmdstr)

# Reconstruct its iterations
for i in range(its):
    ptychopy.epiestep()

# Get back the object and probe result
resobj=ptychopy.epieresobj()
resprobe=ptychopy.epieresprobe()

# Clear up resource
ptychopy.epiepost()

\end{lstlisting}

%

\subsection{Batch GPU mode}

For machines with multiple GPUs, ptychopy provides the feature to run reconstruction jobs on multiple GPUs in two ways. As shown in Fig.~\ref{fig:gui}, the list box provides a list of GPUs that the destination machine has, and the reconstruction job will be put into a queue which is associated with each chosen GPU in the GUI. And then the GPU will use a first-in first-out order to process the assigned job as shown in Fig.~\ref{fig:flow}. This one-to-one GPU allocation is very useful for a large number of jobs with small data size (one GPU can handle one scan). In addition, the job could also run on multiple GPUs and scans with data size larger than one GPU memory would benefit from this. Either way, the controller uses both a worker thread and main thread for controlling the job allocation, running, and visualization. For the first way, the number of worker threads depends on the number of GPUs chosen in the GUI for reconstruction. For each GPU, the controller will create a worker thread for the specified GPU and a separate job queue. For the second way, the reconstruction job will be all put into one thread and use all the GPUs available. The worker threads will record all the job running information and are responsible for triggering the backend CPython module for the reconstruction whenever there are still jobs left in the queue for reconstruction. 

\subsection{Logging and feedback}

Ptychopy has a logging feature to save the reconstruction procedure and the reconstruction parameter information in real-time or after analysis. The information will be redirected to STDOUT and saved under the log folder as ptychopy.log. After the reconstruction, the parameter information and the reconstructed result images could be sent to the email account that is specified on the GUI email address text box. The result image is converted to 8-bit TIFF format in advance, since the resulted reconstruction image could be more than the limited size for the email attachment.






\section{Conclusion}

In this paper we presented a software framework ptychopy which includes a visualization module and a reconstruction module based on ePIE, DM and LSQML algorithms. The ePIE, DM, LSQML phase retrieval algorithms are implemented with CUDA-C for high-performance computing speed-up. This software framework can be flexibly used in ptychographic data reconstruction either with a GUI or in the python script. To support a streamlined workflow for ptychographic imaging processing, the software framework provides a python interface so that it can be easily integrated with beamline control, data collection, and analysis packages. The software has been used in successful operation at the APS, and is easy to be transferred to other ptychography beamlines and synchrotron facilities. 




\section{Acknowledgements}

We would like to thank Olga Antipova for providing the ptychographic reconstructed image from the scan of her sample using our software framework ptychopy. The screenshot for the ptychopy software framework used in this paper were made on a Linux desktop. This work is supported by U.S. Department of Energy, Office of Science, under Contract No. DE-AC02-06CH11357. And the GPU data analysis is tested using Argonne ALCF clusters Cooley and Thetagpu.



\bibliography{reference} 

\begin{thebibliography}{10}

\bibitem{maiden2009improved}
Maiden, A.~M. and Rodenburg, J.~M., ``An improved ptychographical phase
  retrieval algorithm for diffractive imaging,'' {\em Ultramicroscopy}~{\bf
  109}(10),  1256--1262 (2009).

\bibitem{thibault2009probe}
Thibault, P., Dierolf, M., Bunk, O., Menzel, A., and Pfeiffer, F., ``Probe
  retrieval in ptychographic coherent diffractive imaging,'' {\em
  Ultramicroscopy}~{\bf 109}(4),  338--343 (2009).

\bibitem{thibault2013reconstructing}
Thibault, P. and Menzel, A., ``Reconstructing state mixtures from diffraction
  measurements,'' {\em Nature}~{\bf 494}(7435),  68--71 (2013).

\bibitem{miao1999}
Miao, J., Charalambous, P., Kirz, J., and Sayre, D., ``An extension of the
  methods of {X}-ray crystallography to allow imaging of micron-size
  non-crystalline specimens,'' {\em Nature}~{\bf 400},  342--344 (1999).

\bibitem{hoppe1969beugung}
Hoppe, W., ``Beugung im inhomogenen prim{\"a}rstrahlwellenfeld. i. prinzip
  einer phasenmessung von elektronenbeungungsinterferenzen,'' {\em Acta
  Crystallographica Section A: Crystal Physics, Diffraction, Theoretical and
  General Crystallography}~{\bf 25}(4),  495--501 (1969).

\bibitem{faulkner2004movable}
Faulkner, H. and Rodenburg, J., ``Movable aperture lensless transmission
  microscopy: a novel phase retrieval algorithm,'' {\em Physical review
  letters}~{\bf 93}(2),  023903 (2004).

\bibitem{rodenburg2007}
Rodenburg, J., Hurst, A., Cullis, A., Dobson, B., Pfeiffer, F., Bunk, O.,
  David, C., Jefimovs, K., and Johnson, I., ``Hard-x-ray lensless imaging of
  extended objects,'' {\em Physical review letters}~{\bf 98},  034801 (2007).

\bibitem{thibault2008high}
Thibault, P., Dierolf, M., Menzel, A., Bunk, O., David, C., and Pfeiffer, F.,
  ``High-resolution scanning x-ray diffraction microscopy,'' {\em Science}~{\bf
  321}(5887),  379--382 (2008).

\bibitem{luke2004relaxed}
Luke, D.~R., ``Relaxed averaged alternating reflections for diffraction
  imaging,'' {\em Inverse problems}~{\bf 21}(1),  37 (2004).

\bibitem{fienup1982phase}
Fienup, J.~R., ``Phase retrieval algorithms: a comparison,'' {\em Applied
  optics}~{\bf 21}(15),  2758--2769 (1982).

\bibitem{thibault2012maximum}
Thibault, P. and Guizar-Sicairos, M., ``Maximum-likelihood refinement for
  coherent diffractive imaging,'' {\em New Journal of Physics}~{\bf 14}(6),
  063004 (2012).

\bibitem{rodenburg2004phase}
Rodenburg, J.~M. and Faulkner, H.~M., ``A phase retrieval algorithm for
  shifting illumination,'' {\em Applied physics letters}~{\bf 85}(20),
  4795--4797 (2004).

\bibitem{pharr2005gpu}
Pharr, M. and Fernando, R.,  [{\em Gpu gems 2: programming techniques for
  high-performance graphics and general-purpose
  computation}{\nolinebreak\hspace{0.1em}]}, Addison-Wesley Professional
  (2005).

\bibitem{yue2015accelerating}
Yue, K., Nicholas, S., et~al., ``Accelerating laue depth reconstruction
  algorithm with cuda,'' in [{\em Cluster Computing (CLUSTER), 2015 IEEE
  International Conference on}{\nolinebreak\hspace{0.1em}]},   492--493, IEEE
  (2015).

\bibitem{marchesini2016sharp}
Marchesini, S., Krishnan, H., Daurer, B.~J., Shapiro, D.~A., Perciano, T.,
  Sethian, J.~A., and Maia, F.~R., ``Sharp: a distributed gpu-based
  ptychographic solver,'' {\em Journal of Applied Crystallography}~{\bf 49}(4),
   1245--1252 (2016).

\bibitem{nashed2014parallel}
Nashed, Y.~S., Vine, D.~J., Peterka, T., Deng, J., Ross, R., and Jacobsen, C.,
  ``Parallel ptychographic reconstruction,'' {\em Optics express}~{\bf 22}(26),
   32082--32097 (2014).

\bibitem{enders2016computational}
Enders, B. and Thibault, P., ``A computational framework for ptychographic
  reconstructions,'' {\em Proc. R. Soc. A}~{\bf 472}(2196),  20160640 (2016).

\bibitem{odstrcil_oe_18}
Odstr\v{c}il, M., Menzel, A., and Guizar-Sicairos, M., ``Iterative
  least-squares solver for generalized maximum-likelihood ptychography,'' {\em
  Opt. Express}~{\bf 26}(3),  3108--3123 (2018).

\bibitem{daurer2017nanosurveyor}
Daurer, B.~J., Krishnan, H., Perciano, T., Maia, F.~R., Shapiro, D.~A.,
  Sethian, J.~A., and Marchesini, S., ``Nanosurveyor: a framework for real-time
  data processing,'' {\em Advanced structural and chemical imaging}~{\bf 3}(1),
   7 (2017).

\bibitem{radicci_joi_2012}
Radicci, V., Bergamaschi, A., Dinapoli, R., Greiffenberg, D., Henrich, B.,
  Johnson, I., Mozzanica, A., Schmitt, B., and Shi, X., ``Eiger a new single
  photon counting detector for x-ray applications: performance of the chip,''
  {\em Journal of Instrumentation}~{\bf 7}(02),  C02019 (2012).

\bibitem{deng2015continuous}
Deng, J., Nashed, Y.~S., Chen, S., Phillips, N.~W., Peterka, T., Ross, R.,
  Vogt, S., Jacobsen, C., and Vine, D.~J., ``Continuous motion scan
  ptychography: characterization for increased speed in coherent x-ray
  imaging,'' {\em Optics express}~{\bf 23}(5),  5438--5451 (2015).

\bibitem{pelz_apl_2014}
Pelz, P.~M., {Guizar-Sicairos}, M., Thibault, P., Johnson, I., Holler, M., and
  Menzel, A., ``On-the-fly scans for x-ray ptychography,'' {\em Applied Physics
  Letters}~{\bf 105},  251101 (2014).

\bibitem{huang_scirep_2015}
Huang, X., Lauer, K., Clark, J.~N., Xu, W., Nazaretski, E., Harder, R.,
  Robinson, I.~K., and Chu, Y.~S., ``Fly-scan ptychography,'' {\em Scientific
  Reports}~{\bf 5},  9074 (2015).

\bibitem{deng2019velociprobe}
Deng, J., Preissner, C., Klug, J.~A., Mashrafi, S., Roehrig, C., Jiang, Y.,
  Yao, Y., Wojcik, M., Wyman, M.~D., Vine, D., et~al., ``The velociprobe: An
  ultrafast hard x-ray nanoprobe for high-resolution ptychographic imaging,''
  {\em Review of Scientific Instruments}  (2019).

\bibitem{newville2014pyepics}
Newville, M., ``Pyepics, epics channel access for python,'' (2014).

\bibitem{ptychopy}
Yue, K., ``https://github.com/kyuepublic/ptychopy,'' (2019).

\end{thebibliography}
\bibliographystyle{spiebib} 

\end{document}